# MIP—An AI Distributed Architectural Model to Introduce Cognitive computing capabilities in Cyber-Physical Systems (CPS)

**Authors:**
Pasquale Giampa <p.giampa@redev.technology>
    (Head of Infrastrucutre - Redev Technology Ltd)
Massimiliano Dibitonto <m.dibitonto@unilink.it>
    (Head of Data Science - Redev Technology Ltd, Researcher - Link Campus University – https://unilink.it)

**Abstract** This paper introduces the MIP Platform architecture model, a novel AI-based cognitive computing platform architecture. The goal of the proposed application of MIP is to reduce the implementation burden for the usage of AI algorithms applied to cognitive computing and fluent HMI interactions within the manufacturing process in a cyber-physical production system. The cognitive inferencing engine of MIP is a deterministic cognitive module that processes declarative goals, identifies "Intents" and "Entities", selects suitable actions and associated algorithms, and invokes for the execution a processing logic ("Function") configured in the internal Function-as-a-Service or Connectivity Engine.
Constant observation and evaluation against performance criteria assess the performance of Lambda(s) for many and varying scenarios.
The modular design with well-defined interfaces enables the reusability and extensibility of FaaS components.
An integrated BigData platform implements this modular design supported by technologies such as Docker, Kubernetes for virtualization and orchestration of the individual components and their communication. The implementation of the architecture is evaluated using a real-world use case later discussed in this paper.

Artificial Intelligence
Machine Learning
Cognitive Computing
Reference Architecture
Cognition
Big Data Platform
Modularization
Docker
Kubernetes

## Introduction

The adoption of Artificial Intelligence (AI) and Machine Learning (ML) technologies in the manufacturing industry can help to significantly reduce costs and provides new market opportunities. Originally referred to as AI, academics in the '90s began to use the term "Cognitive Computing" instead, to refer to that science that was designed to teach computers to think like a human mind.

Cognitive computing refers to the development of software designed after the human brain, which has NLP (Natural Language Processing) and NLU (Natural Language Understanding) capabilities, learns from experience, interacts with humans through spoken language, and supports decision making based on experience. Their way of processing conversations is canonically neither linear nor deterministic.

Cognitive Computing software effectively integrates technology and biology in a novel attempt to re-engineer the human brains, the most efficient and effective computers on Earth.

The application of Cognitive Computing technologies to control Cyber-physical Production Systems (CPPS) [1] marks a further step towards the complete automation of the shop floor of our industries.

Many Industrial-IoT applications rely on the use of Artificial Intelligence or Machine Learning [2] technologies to enable use-cases like:

- work order optimization,
- condition monitoring,
- preventive or predictive maintenance,
- anomaly detection,
- optimization.

The lack of standards, market fragmentation and diversity of problems to be solved in Manufacturing, makes the adoption of AI and ML technologies vastly inefficient on the vast majority of the potential target market (SMEs) [3].

Our goals (G) for the application of the MIP architectural model are represented by:

- (G-1): **Reliability**: In an industrial control use-case, determinism of control is a mandatory constraint for the applicability of the overall solution. Since MIP is by design a distributed, micro-service based architecture, automatically monitored, scalable and also redundant by design, using the at-least-once [4] delivery pattern, message losses or downtime, are extremely unlikely, making the whole control system stable and reliable.
- (G-2): **Flexibility**: A significant drawback of existing architectures is the need of adding new "platform-level" code to extend the platform capabilities. This inflexibility is not acceptable in a CPPS scenario. The MIP has to be easily extendable to enable quick and reliable implementations of new features and integrations.
- (G-3): **Generalizability**: The MIP has to be generic enough to support many different interactions use cases. Thus, the MIP has to be able to process user "inputs" from different "channels" like Voice (human), Text (human), API (AI-driven) method invocation, has to understand the intent and invoking the specific "outcome" function through the *FaaS (Function-as-a-Service) Engine*, and learn over time from external systems to improve the system's performance acquiring new data, in particular from "machines" and this is enabled by the *Connectivity Service*.

Analysing the already existing architectures associated with our goals (G1-G3), at the best of our knowledge, no work tackles those goals effectively. Architectures analysed like the Industrial Internet Reference
Architecture (IIRA) [5], or the 5C architecture [1], do not cover at the required level of abstraction ignoring completely tangible architecture details. They can't immediately help addressing implementation aspects of real industrial use cases.

Since there is an intersection between our architecture and these methods, they are considered for our implementation.

Big Data Platforms (BDPs) to the best of our knowledge do not fulfil by themselves cognitive architecture needs since the obvious lag while they perform efficiently on non-real-time jobs at scale [6]. This doesn't mean they aren't a critical element of a larger architecture as they actually are in MIP.

About the context of Industry 4.0, in the scope of CAAI [7] **cognition** refers to "all processes by which the input data is transformed, reduced, elaborated, stored, recovered, and used to solve I4.0 use cases, i.e., condition monitoring, anomaly detection, optimization, and predictive maintenance" [7].
Following our analysis and goals, **cognition** refers to "all processes by which the human control commands, once issued are securely, reliably and univocally referred to a single or more actuation on the CPPS".

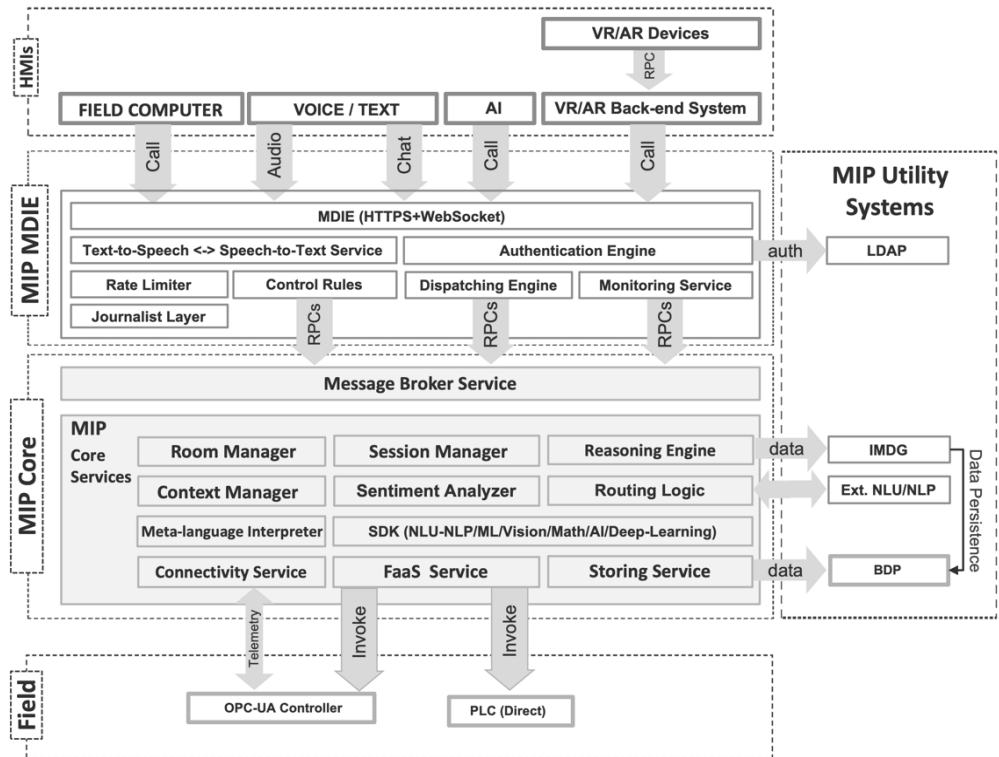

*Figure 1: MIP Overall Architecture*

The contribution of this paper is a novel cognitive computing architecture to provide deterministic control and querying for CPPS to humans through current spoken language (our observation on this paper are focused exclusively on the English spoken language).
The MIP architecture has several advantages in comparison to other the state-of-the-art architectures.
To tackle goals (G-1) to (G-3), the following methods (M) are considered in this paper:

- (M-1): **Multichannel approach**: Transparent enablement of a seamless integration of multiple "interaction channels" (for both human and AI-driven inputs) is ensured by the MIP MDIE (Multi-channel Data Ingestion Engine). The MDIE internal architecture is divided into two different layers. A first outer layer that provides "protocol translation" capabilities to translate any input protocol (either text itself, voice, binary protocols, etc) into a their text representation and further inner layer in charge of converting these *text-only* datagrams into a common "Meta-language" later ingested and decoded by the internals of the MIP Core platform portion.
- (M-2): **FaaS (Function-as-a-Service) Engine**: the ease of extensibility without compromising the product codebases enable effectively the *generalizability* (G-3) of the MIP. Extensibility is achieved integrating new features in terms of triggered actions, language processing extensions, custom rules, etc, by adding new "Lambda Functions" through the FaaS Engine.
- (M-3): **At-least-Once approach**: A design goal of MIP is to be reliable in every single component of its infrastructure to enable the reliability (G-1) goal. To achieve such goal to adapt MIP to even the most concerning and critical use-cases, it has been designed following a microservice-oriented architecture, with a central message broker used to connect all the microservices between them allowing the exchange of data and RPC (Remote Procedure Call) execution.

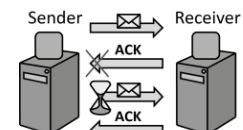

*Figure 2: At-least-Once model*

To make it deeply reliable the underlying message broker has been configured in "at-least-once" mode [4]. At-least-once provides a low-level (broker-level) delivery guarantee of every single message. It is less efficient in terms of amount of broker traffic than a "at-most-once" model, but in our case the input volumes are way below the threshold of prioritizing a "at-most-once" model against a "at-lease-once" one.

**The MIP Architecture**
In this section, we introduce the MIP architecture.

*Overview*
A cognitive architecture like MIP designed to fit the task of controlling a CPPS builds upon the idea of modelling inputs and streams taken in input to achieve the goals (G-1,2,3).
The architecture, depicted in Figure 1, is based on a multi-tier architecture to ensure horizontal scalability and reliability.
The MIP Core Services layers wraps a number of engines, here are the ones relevant to this paper's goals:

- **Session Manager**: Gives extra-conversational memory to the MIP platform, "allowing humans to ditch" and later "pick-up" the conversation with the same context information and state variables.
- **Routing Logic layer**: On a multi-AI use-case allows MIP to decide which AI to invoke for the next NLU/NLP request through a pre-established ruleset. Whether the internal one or another external engine,

- Routing Logic layer acts promptly for the correct routing.
- **Context Manager**: Gives intra-conversational memory to the MIP platform, allowing to switch "entity" while keeping the dialog flow on the same intent, avoiding traversing again the intents tree.
- **Meta-Language Interpreter**: is in charge of decoding the Meta-language generated by the MDIE engine, breaking it down into a sequence of cognitive micro-operations later ingested by the Reasoning Engine.
- **Reasoning Engine:** is the core service among the MIP Core Service. The Reasoning Engine decodes the cognitive micro-operations received by the Meta-Language Interpreter to generate events that can trigger FaaS functions. Decisions are stored on an IMDG (In-Memory Data Grid) and the Reasoning Engine is in charge of querying the pre-defined ruleset(s) saved on the IMDG. There are two FaaS functions that are registered in the MIP as "system function": the "answering logic" FaaS and the "HTTP Rest" FaaS. The first one provides the capability to return the "answer" always to the same channel where it came from while the second one allows to program a generic HTTP Rest request and to bind it to a specific cognitive microoperation.
- **Message Broker:** it is the core communication infrastructure of MIP. It provides asynchronous, "at-least-once" communications to all the software modules of MIP. All the modules implement a request-response paradigm, through the adoption of publisher / subscriber mechanism [8] that allows easy addressing of who is making the request (publisher), how to reach who is processing the request (subscriber), how to return the answer to the publisher.
- **Authentication Engine:** provides to every component of the MIP platform the "authentication" and "authorization" capabilities. The Authentication Engine itself connects to an LDAP [9] server.
- **Connectivity Service:** provides further support to the *FaaS Engine* to ease the integration of data coming from machines. The MIP Connectivity Service is a component of the overall MIP Core Services middleware that provides an abstraction layer for the implementation of bidirectional protocol translation layers towards devices speaking different protocols (a.k.a. "Protocol Mediation") which data requires to be acquired and exported from and to the external device. A typical example is the need to implement communications towards devices speaking TCP protocols that requires on or more persistent connections vs an event driven connection model (supported via the MIP FaaS Engine).

*Scalability*

With the eventual usage of automated control and querying capabilities of the CPPS through the AI/API of MIP, it is inevitable that the virtual machines running underneath the platform will reach a saturation point somewhere in the processing chain, highlighting the known limitations of "non-distributed architectures" vs "distributed architectures". Non distributed architectures reach the normal scalability bottleneck when the compute load is concentrated by design on a single machine and there is by design no vehicle to distribute the load on more than 1 node, generating a scalability lock when the CPU(s) of the machine handling a specific software module of the platform hits 100% of any resource (CPU, RAM, Network, IO, etc).

MIP has been designed as a "distributed architecture" [9] leveraging on a messaging system (the "Message Broker"), itself another distributed and indeed scalable component.

All the different modules communicate each other using the broker, providing a reliable, fast data exchange bus between the MIP services.

MIP shows to be a pretty fit for the job since achieves the specific reliability required to satisfy our first goal, (G-1). This goal is achieved mainly through a specific capability of the Message Broker named "Consumer Groups" or "Shared Subscriptions" [10].

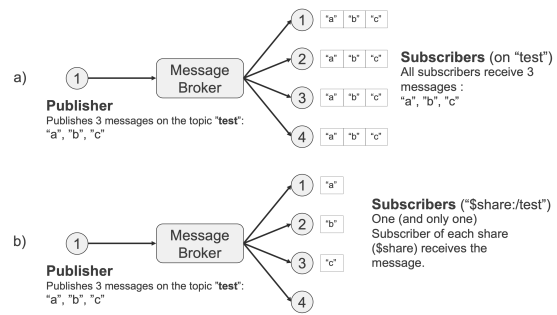

*Figure 3: Consumer Groups / Shared Subscription working model*

This mechanism behind the Shared Subscription model is very simple and effective to guarantee scalability and reliability of a distributed system.

With standard the subscribe mechanism every subscriber gets its own copy of every message sent by the published which matches the subscribed topic (as we can see in a) of figure 3).

When using shared subscriptions, each subscription group ($share in the example of figure 3), which can be virtually seen as a virtual client, acts as proxy for multiple real subscribers. The MIP Message Broker then selects one subscriber of the group and delivers the message (like seen in b) of figure 3). Shared Subscriptions are designed to relieve cluster traffic and latency dramatically for high scalability deployments. In fact, *Shared Subscriptions are the recommended way to connect horizontally scaling backend systems with last generation message brokers*.

In the diagram in figure 4 we can see for each case of figure 2, how different subscribers positioned on different cluster nodes (different container instances / virtual machines) behave both when using or not Shared subscriptions.

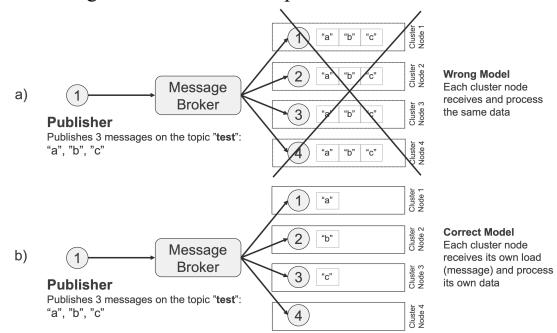

*Figure 4: Multi-node Consumer Groups / Shared Subscriptions*

*Disposable Architectural Components*

Big SaaS software companies have proved to the world that it's easier and more cost effective to assume an architectural component will fail and instead design architectures that can cope with failures. There is one easy reason, everyone wants to buy compute resources as cheaply as possible, determining that some parts of its infrastructure may run on cloud resources that aren't guaranteed and eventually on different cloud providers.

Some cloud resources used by the platform could be torn down at any time if resources are needed by the cloud provider to service a customer willing to pay more (i.e. AWS EC2 Spot Instances [12]). When these compute units are torn down, the company doesn't want its platform to provide a poor experience to the end user, so the infrastructure must be able to cope by design with the loss of these compute unites (mostly containers and virtual machines) without a loss of service.

This smart architectural model (as described in b) of the figure above) other than providing horizontal scalability gives out of the box the ability to enable MIP to embrace the ability to claim itself a platform with "disposable architectural components". Any component of the platform marries this architectural model and doesn't need to rely on an extremely stable and expensive virtual machine / container infrastructure. Deploying the MIP platform on AWS Spot instances or on old hardware doesn't imply (as long as the Kubernetes cluster has enough nodes) that the platform is less stable to the end user against a deployment on a more expensive infrastructure.

*Ease of Integration and Generalizability*
As already mentioned in (M-2) the MIP *Connectivity Service* and the MIP *FaaS Engine* provide ease of extensibility without compromising the product codebase and its reliability.

Before going deeper specifically into the FaaS Engine there is another architectural term that needs to be introduced and that is worth more than just a mention: "Serverless Computing".

Serverless computing is a computing model which aims to abstract server scalability and low-level infrastructure management away from developers. This model implies that allocation of resources is managed by the underlying *virtualization layer*, instead of the application developer, providing huge benefits to the development lifecycle. In other words, serverless aims to do exactly what it sounds like — allow applications to be developed without concerns about implementing, tweaking, or scaling a server (at least, to the perspective of a developer).

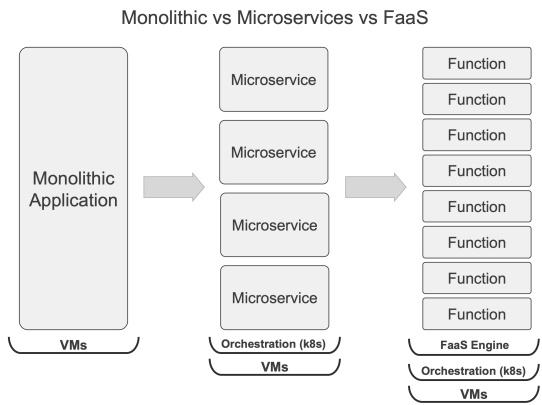

*Figure 5: FaaS / Serverless model*

What this means is that developers can simply upload modular chunks of functionality into the MIP FaaS Engine that are executed independently and orchestrated and scaled automatically accordingly to events and request volumes.

New features to integrate external components, for both in outbound and inbound data flow, are easily implemented as new "Functions" (a.k.a. "Lambdas") and are brought up, invoked, kept alive, orchestrated (at scale) and finally terminated by the FaaS Engine. The FaaS Engine itself runs on top of Kubernetes that provides multi-node horizontal scalability.

*BigData Platform*
The MIP BDP (BigData Platform) is a distributed platform based on Hadoop and HDFS (the Hadoop Distributed Filesystem) plus a plethora of microservices and utility daemons. Let's focus on the most important aspects of it:
- **Data Storage (HDFS):** HDFS is a distributed file system designed to run on commodity hardware, it is highly fault-tolerant and is designed to be deployed on low-cost hardware. HDFS provides high throughput access to application data and is suitable for applications that get to have very large data sets. Through its client interface, Hadoop allows to manage files. Files are divided into blocks, and file access follows *multi-reader*, *single-writer* semantics. To achieve the fault-tolerance multiple *replicas* of a block are stored on different *DataNodes*. The number of replicas is called the *replication factor*. When a file is accessed for *append* the HDFS write operation creates a pipeline of DataNodes to receive and store the replicas. Following writes to that block go through this pipeline. For reading operations HDFS assigns to the client one of the DataNodes holding copies of the block and request a "read" from it.

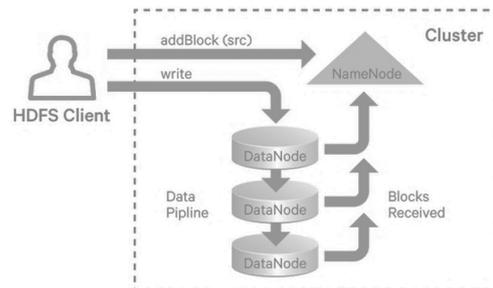

*Figure 6: HDFS Data Pipeline model*

- **Virtualization and Containerization**: This applies to the whole MIP platform and is not exclusive to the BigData Platform. All the software components of the MIP platform exist as instances of the related Docker Containers images. Containers allow the isolation of the single module components from the general environment, ensuring that all the computing requirements for the lifecycle of a specific module are met and that single software modules do not interfere each other on the same platform, similar to what happens with virtual machines. As opposite to a virtual machine, a container uses the host operating system and kernel and shares binaries, libraries and networking, which results always in less overhead. Containers use *layers* to create differential updates from the starting point of the container. They are consistent and immutable, which ensures compatibility across orchestration platforms. MIP provides A container registry store that hosts the container images and keeps track of changes via versioning. Docker is the container engine for the implementation of this use-case.
- **Orchestration**: The MIP platform manages (instantiate, destroy, keeps running and destroy) containers to orchestrate the necessary platform components to deal with increased or decreased load. MIP orchestration is based on Kubernetes (shortened K8S). K8s handles deployments, configuration, updating, and removing of the virtualized software components. A configuration file declaratively composes the whole system and lists the different services.
- **Microservices**: All the MIP engines and services are developed as microservices. Each one of them includes standardized communication libraries (that gets updated harmonically throughout updates in every single microservice. These communication libraries support functionalities to publish and subscribe to relevant topics on the MIP Message Broker. This approach makes the whole platform modular, language-agnostic (each microservice can be written in any language), and utilizes well-defined data model and data format (JSON). Following microservice best-practices, microservices can store internal data in their local storage even though, in case of MIP most of them are stateless.

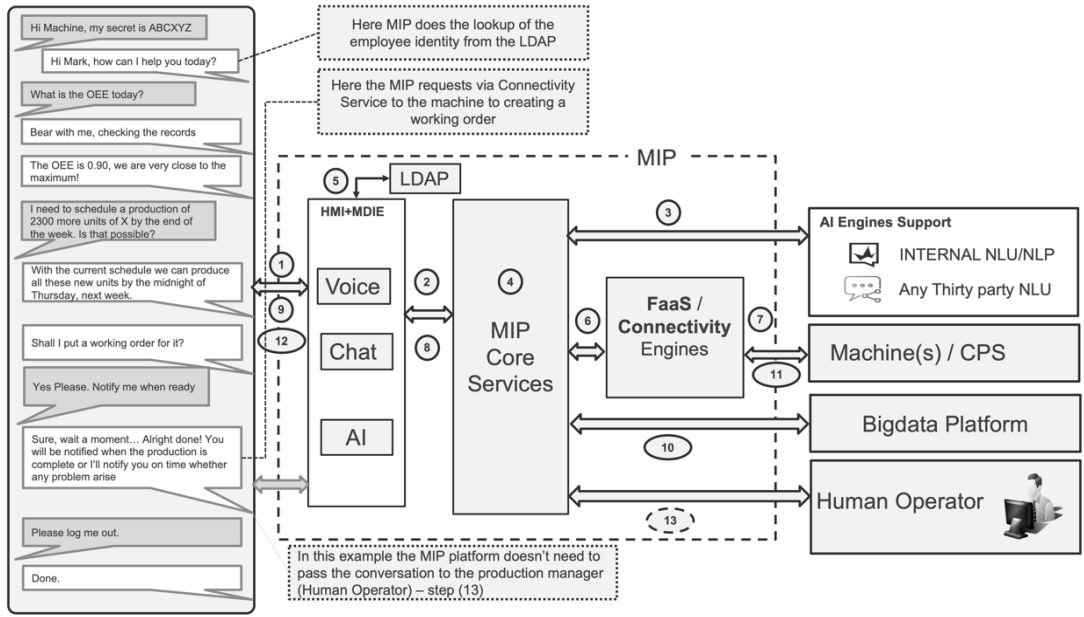

*Figure 7: MIP interaction process*

*Use-Case:*
The scope of this paper is to describe the architecture underneath the hood of the MIP platform, meant as a reliable, distributed Cognitive Computing architecture to provide a smart, reliable and extensible control tool for the Cyber-Physical system. We introduce briefly the process behind a very simple and straightforward, end-to-end control use-case of a machine (CPS) connected to the MIP platform.

For sake of simplicity of this on-purpose purely architectural paper, further details about the internals of the AI of this platform are demanded to future scripts.

This paragraph wants to highlight the end-to-end process of the MIP Platform acting as a control Voice control system for the machine. The process above is a simplified version of the real process and is divided into 13 macro-steps:

- (S1): The user is "talking" to the "machine" through a web interface providing voice-recording and transmission capabilities via WebRTC [13]. Voice reaches the MDIE platform portion where audio data is ingested (after the *Rate Limiter* has given it permission to pass through) and passed to the Speech-to-Text Engine the one in charge of translating audio to text. After the conversion, text is passed in sequence to Journalist that decides the "dispatching" data model following the source that has provided the audio (now text).
- (S2): Once the finalized object (now a JSON) is ready, it is passed to the *Dispatching Engine* that provides a reliable publishing client against the *Message Broker*.
- (S3): Once the text has reached the MIP Core Services, a dialog room is created by the *Room Manager* and a session is generated for the current conversation. A session ID is integrated in the JSON message and it is passed to the integrated (or external) NLU/NLP service, that is in charge of telling us: The *Intent* and the *Entity*. To provide an easy example of what Intent and Entities are, let's consider the fast phrase spoken by the human to the machine "Hi Machine, my secret is ABCXYZ". With This phrase the NLU/NLP moves its *context* to the #LOGIN intent (for sure since the "secret" word has been seen in the sentence) and applies as entity #MACHINE. This is pretty straightforward to be understood since we're actually *trying to authenticate ourselves with the machine*.
- (S4): #LOGIN and #MACHINE along with the #SECRET extracted from sentence are passed back from the NLU/NLP service straight on the "authentication" topic of the broker, where the *Authentication Engine* is continuously waiting for a request.
- (S5): The request is received by the Authentication Engine that dispatches a query towards the internal (or external) LDAP server where all the credentials, roles and access control lists are stored. Upon successful authentication, the Authentication Engine (that's keeping track of the request sends a welcome message to the user.
- (S6): At this point the user makes another request that goes again through (S1) to (S6) (skipping S5 since the Authentication Engine has now stored an *AccessToken* into the platform internal In-Memory Data Grid (IMDG) to provide fast re-authorization and re-authentication until a *logout request* arrives or until such token expires. The new user request goes all the way through the "decision engines" of the MIP Core Services, leading to the generation of a "Request" for the machine, to obtain the "OEE" (OEE stands for Overall Equipment Effectiveness) [14] indicator.
- (S7): The request for the machine is passed via an existing OPC-UA connection between the OPC-UA Server of the machine and the *Connectivity Service* of the MIP Platform.
- (S8): A specific variable for the OEE is read via OPC-UA and the result obtained is passed back to the MDIE via the MIP Core Services
- (S9): Finally, this "value" that has been enriched with a context message, is converted to voice using the Text-to-Speech engine of the MDIE.
- (S10): At the same time that the message is delivered, the whole session information, including request, response, intent and entity are stored into the MIP *BigData Platform* for future usage.
- (S11): The user once received the answer about the OEE by the MIP, decides to instruct the machine to activate a new working order for further 2300 units by the end of the following week. This new message goes again through (S1)→ (S6) and finally an order is passed to the machine.
- (S12): A confirmation answer is passed back to the user, following again S8 and S9.
- (S13): A step is avoided, the nr. 13, since the whole process has been correctly managed by the AI and no interaction with a human operator has resulted required.

*Results*
This whole process for this paper has been tested in lab, end-to-end, on the real use-case reported in it. A real OPC-UA

connection to an OPC-UA server has been created to simulate a real request happening to read the OEE and to dispatch a working order.

A simulation of a "non-dispatchable" working order has been tested, [4] both increasing the number of units in the timeframe requested or reducing dramatically the OEE read by the UPA-UA Server.

The platform has been installed on a pair of Dell PowerEdge servers, both with the same hardware and software specs:

- *RAM*: 64Gb DDR4
- *Storage*: 3 TB of storage provided by multiple SSD Disks
- *CPU*: Dual Xeon 20 cores Xeon CPUs
- *Virtualization*: VMWARE ESXI 6.5
- *Orchestration*: Kubernetes 1.17
- Containers: All container images are based on Alpine Linux

*Conclusion and Outlook*

In this paper, we defined goals (G1-G3) to be reached by the use of the MIP Platform. Each goal is addressed by a particular method (M1-M3), which is be implemented by several technologies, solutions or platform components related to the MIP Platform.

Besides the main contribution of this paper, the proposed AI Cognitive Architecture, the major results (R) can be summarized as follows:

- (R-1): The main objective, to provide a design of a cognitive computing architecture to create a new generation of HMIs for CPPS control tasks, cannot be reached by addressing single goals or implementing single methods. The main implication is that only considering both the whole architecture along with its particular configuration (to mention a few: the At-least-once model, and the Shared Subscriptions model) and NLU/NLP training, can provide an MVP (Minimum Viable Product) for a real world, voice actioned digital assistant able to control a CPPS.
- (R-2): The defined goals and methods are a reference character of MIP architecture and will keep being a valid model over a long time. However, the solutions and chosen technologies, configurations or concepts are subject to change in the future. A simple example is that Docker is a valid choice to fulfil the requirements of this paper but as technologies move forward, this option may be replaced by a more modern one even though can be considered as a state-of-the-art virtualization method nowadays.
- (R-3): The performance of a given algorithm or neural-network used for a given use case may change over time (worsening and improving), therefore the whole system has to integrate a fully integrated benchmarking suite able to prove over the time the platform intelligence compliance with the given task.
- (R-4): From our point of view, these outcomes altogether cover three important disciplines, including related applications and methods:
  - **Infrastructure**: best practice to develop package, deploy and run a modern platform (Containers, VMs)
  - **BigData**: working principles of the core of all BigData platforms (Hadoop/HDFS)
  - **Distributed Systems**: best practices and tools to design a truly horizontally scalable application
  - **Cognitive Platforms**: best practices to use a cognitive computing framework
  - **Distributed Data Processing**: best practices to develop microservices that can actually scale while processing increasing amounts of data
  - **Industrial IoT**: principles of integration of CPPS and industrial controllers (through the adoption of the fastest growing standard protocol for the industries – OPC-UA)

The work on this paper brings up some questions, leading to logical next steps and follow-up research tasks:

- Possible interesting use-cases for self-controlled production schedules.
- Implementation of an automated remote management service for spare parts of the machine.
- Studying the overall design of an End-to-end automated business platform, from the purchase of. A good on an e-commerce, to the generation of sales order, the automated manufacturing of the good, its delivery to the end-user.

**Conflict of interest**

The authors declare that they have no conflict of interest.